\documentstyle[12pt]{article}
\advance\textheight by \topskip
\begin{document}
\thispagestyle{empty}
\begin{flushright}
SU--ITP--95--11\\
astro-ph/9506017\\
June 2, 1995\\
\end{flushright}
\vskip 2 cm
\begin{center}
{\LARGE\bf Inflation  with   $\Omega \not = 1$}
\vskip 1.7cm

 {\bf Andrei Linde}\footnote{
E-mail: linde@physics.stanford.edu} \ {\bf and Arthur
Mezhlumian}\footnote{E-mail: arthur@physics.stanford.edu}
\vskip 1.5mm
Department of Physics, Stanford University,
Stanford, CA 94305--4060, USA
\end{center}
\vskip 2cm

{\centerline{\large\bf Abstract}}
\begin{quotation}
\vskip -0.4cm
 We  discuss various models of inflationary    universe with $\Omega \not = 1$.
A homogeneous  universe with $\Omega > 1$ may appear due to creation of the
universe ``from nothing'' in the theories where the effective potential becomes
very steep at large $\phi$, or in the theories where the inflaton field $\phi$
nonminimally couples to gravity. Inflation with $\Omega < 1$ generally requires
intermediate first order phase transition with the bubble formation, and with a
second stage of inflation inside the bubble. It is possible to realize this
scenario in the context of a theory of one scalar field, but typically it
requires artificially bent effective potentials and/or nonminimal kinetic
terms. It is much easier to obtain an open universe in the models involving two
scalar fields.   However, these models   have their own specific problems. We
propose three different   models of this type which can describe an open
homogeneous inflationary universe.

\end{quotation}
 \newpage

\baselineskip=16pt

\section{Introduction}
One of the most robust predictions of inflationary cosmology is that the
universe after inflation becomes extremely flat, which corresponds to $\Omega =
1$. Here $\Omega = {\rho\over \rho_c}$,\,  $\rho_c$ being the energy density of
a
flat universe.  There were many good reasons to believe that this prediction
was quite generic.  The only way to avoid this conclusion is to assume that the
universe inflated only by about $e^{60}$ times. Exact value of the number of
e-foldings $N$ depends on details of the theory and may somewhat differ from
60. It is important, however, that in any particular theory  inflation by an
extra factor $e^2$ would make  the universe with $\Omega = 0.5$ or with
$\Omega = 1.5$ almost exactly flat. Meanwhile, the typical number of
e-foldings, say, in chaotic inflation scenario in the theory ${m^2\over 2}
\phi^2$ is not 60 but rather $10^{12}$ \cite{MyBook}.

One can construct models where
inflation leads to expansion of the universe by the factor $e^{60}$. However,
in most of such models small number of e-foldings simultaneously implies that
  density perturbations
are extremely large.  Indeed, in most inflationary models the process of
inflation begins at the point where density perturbations
${\delta\rho\over\rho}
\sim {H^2\over \dot\phi}$ are very large. The simplest example is the original
new inflation scenario \cite{New}, where inflation begins at the top of the
effective potential
with $\dot\phi = 0$. If there are less than $60$ e-foldings from this moment to
the end of inflation, then we would see extremely large density perturbations
on the scale of horizon.

It may be possible to overcome   this obstacle by a specific choice of the
effective potential. However, this would be only a partial solution. Indeed, if
the universe
does not inflate long enough to become flat, then by the same token it
does not inflate long enough to become homogeneous and isotropic.
Thus,   the main reason why it is difficult to construct inflationary models
with $\Omega \not = 1$ is not the issue of fine tuning of the parameters of the
models, which is necessary to obtain the universe inflating exactly $e^{60}$
times, but the problem of obtaining a homogeneous universe after inflation.

Fortunately, it is possible to solve this problem, both for a closed universe
\cite{Lab,Omega} and for an open one \cite{Omega}--\cite{BGT}. The
main idea is to use the well known fact that the region of space created in the
process of a quantum tunneling tends to have a spherically symmetric shape,
and homogeneous interior, if the tunneling process is suppressed strongly
enough. Then such bubbles of a new phase  tend  to evolve (expand) in a
spherically symmetric
fashion. Thus, if one
could associate the whole visible part of the universe with an interior of one
such region, one would solve the homogeneity problem, and then all other
problems
will be solved by the subsequent relatively short stage of inflation.

For a closed universe the realization of this program is relatively
straightforward
\cite{Lab,Omega}. One should consider the process of quantum creation of  a
closed inflationary universe from ``nothing.''  If the probability of such a
process is exponentially suppressed (and this is indeed the case if inflation
is possible only at the energy density much smaller than the Planck density
\cite{Creation}), then the universe created that way will be  rather
homogeneous from the very beginning.

The situation with an open universe is much more complicated. Indeed, an open
universe is infinite, and it may seem impossible to create an infinite universe
by a tunneling process. Fortunately, this is not the case: any bubble formed in
the process of the false vacuum decay looks from inside like an infinite open
universe \cite{CL}--\cite{STYY}.
 If this universe continues inflating
inside the bubble \cite{Gott,BGT,Omega}, then we obtain an open inflationary
universe.

These possibilities became a subject of an active investigation only very
recently, and there are still many questions to be addressed. First of all, the
bubbles created by tunneling are not {\it absolutely} uniform even if the
probability of tunneling is very small.   This may easily spoil the whole
scenario since in the
end of the day we need to explain why the microwave background radiation is
isotropic with an accuracy of about $10^{-5}$. Previously we did not care much
about initial homogeneities, but if the stage of inflation is short, we will
the see original
inhomogeneities imprinted in the perturbations of the microwave background
radiation.

The second problem is to construct  realistic inflationary models where all
these ideas could be realized in a natural way. Whereas for the closed universe
this problem can be easily solved \cite{Lab,Omega}, for an open universe we
again meet complications. It would be very nice to to obtain an open universe
in a theory of just one scalar  field \cite{BGT}. However, in practice it is
not very easy to obtain a satisfactory model of this type. Typically one is
forced either to introduce very complicated effective potentials, or consider
theories with nonminimal kinetic terms for the inflaton field  \cite{Bucher}.
This makes the models   not only
fine-tuned, but also rather complicated. It  is very good to know that  the
models of such type in principle can be constructed, but  it is also very
tempting to find a
more natural realization of the   inflationary universe scenario which would
give
inflation with $\Omega < 1$.

 This goal
can be achieved if one considers models of two
scalar fields \cite{Omega}. One of them may be the standard inflaton
field
$\phi$ with a relatively small mass, another may be, e.g., the scalar field
responsible for the symmetry breaking in GUTs. The presence of two scalar
fields allows one to obtain the required bending of the inflaton potential by
simply changing the definition of the inflaton field in the process of
inflation. On the first stage the role of the inflaton is played by a heavy
field with a steep barrier in its potential, while on the second stage the
role of the inflaton is played by a light field, rolling in a flat direction
``orthogonal'' to the direction of quantum tunneling. This change of the
direction of evolution in the space of scalar fields removes the naturalness
constraints for the form of the potential, which are present in the case of one
field.

Inflationary models of this type
are quite simple, yet they have many interesting features. In these models
the universe consists of infinitely many expanding bubbles immersed into
exponentially expanding false vacuum state. Each of these bubbles inside looks
like an open universe, but the values of $\Omega$ in these universes may take
any value from $1$ to $0$.
In some of these models the situation is even more complicated: Interior of
each bubble looks like an infinite  universe with an effective value of
$\Omega$
slowly decreasing to $\Omega = 0$ at an exponentially  large distance from the
center  of the bubble. We will call such universes quasiopen. Thus, rather
unexpectedly, we are obtaining a large  variety of  interesting and previously
unexplored possibilities.

In this paper we will continue our discussion of inflationary models with
$\Omega \not = 1$. In Section 2 we will describe a model  of a closed
inflationary universe. In Section 3 we will consider the possibility to
implement  an open inflation scenario in the theory of one
scalar field.   In Section 4 we discuss the issue of a
spherical symmetry of the bubbles produced by a tunneling process. In
Sections \ref{Simplest} --
\ref{Natural} we will describe several different models of an open inflationary
universe. Finally, in the last Section of the paper we will summarize our
results and discuss the most
important question: What does inflationary cosmology say now to those
trying to determine $\Omega$ from observational data?

\section{\label{Closed} Closed inflationary universe}

For a long time it was not quite clear how can one obtain a homogeneous closed
inflationary universe. In  \cite{BGT} it was even argued that this is
impossible. Fortunately, it turns to be a  relatively easy task
\cite{Lab, Omega}. For
example, one can consider a particular version of the chaotic inflation
scenario
\cite{Chaotic} with the effective potential
\begin{equation}\label{1}
V(\phi) = {m^2 \phi^2\over 2}\, \exp{\Bigl({\phi\over CM_{\rm P}}\Bigr)^2} \ .
\end{equation}
Potentials of a similar type often appear in supergravity. In this theory
inflation occurs only in the interval ${M_{\rm P}\over 2} {\
\lower-1.2pt\vbox{\hbox{\rlap{$<$}\lower5pt\vbox{\hbox{$\sim$}}}}\ } \phi {\
\lower-1.2pt\vbox{\hbox{\rlap{$<$}\lower5pt\vbox{\hbox{$\sim$}}}}\ }
CM_{\rm P}$. The most natural way to realize inflationary scenario in this
theory is
to assume that the universe was created ``from nothing''  with the field $\phi$
in the interval  ${M_{\rm P}\over 2} {\
\lower-1.2pt\vbox{\hbox{\rlap{$<$}\lower5pt\vbox{\hbox{$\sim$}}}}\ } \phi {\
\lower-1.2pt\vbox{\hbox{\rlap{$<$}\lower5pt\vbox{\hbox{$\sim$}}}}\ }
CM_{\rm P}$. The universe at the moment of its creation has a size $H^{-1}$,
and then it begins inflating as $H^{-1} \cosh{Ht}$. According to
\cite{Creation}--\cite{Vilenkin}, the probability of creation of an
inflationary
universe is
suppressed by
\begin{equation}\label{2}
P \sim \exp\Bigl(-{3M^4_{\rm P}\over 8V(\phi)}\Bigr) \ .
\end{equation}
Therefore the maximum of the probability appears near the upper range of values
of the field $\phi$ for which inflation is possible, i.e. at $\phi_0 \sim C
M_{\rm P}$ (see more discussion about this below).
The probability of such an event will be so strongly suppressed that the
universe will be formed almost ideally homogeneous and spherically symmetric.
As  pointed out in   \cite{Lab}, this  solves the homogeneity, isotropy and
horizon problems even before inflation really takes over. Then the size of the
newly born universe in this model expands by the factor $\exp({2\pi
\phi_0^2M_{\rm P}^{-2}})\sim \exp({2\pi C^2})$ during the
stage of inflation \cite{MyBook}. If $C {\
\lower-1.2pt\vbox{\hbox{\rlap{$>$}\lower5pt\vbox{\hbox{$\sim$}}}}\ } 3$, i.e.
if $\phi_0 {\
\lower-1.2pt\vbox{\hbox{\rlap{$>$}\lower5pt\vbox{\hbox{$\sim$}}}}\ } 3M_{\rm P}
\sim 3.6\times 10^{19}$ GeV, the
universe expands  more than $e^{60}$ times, and  it becomes very flat.
Meanwhile, for $C \ll 3$ the universe always remains ``underinflated'' and very
curved, with $\Omega \gg 1$. We emphasize again that in this particular model
``underinflation" does not lead to any problems with homogeneity and isotropy.
The only problem with this model is that in order
to obtain $\Omega$ in the interval between $1$ and $2$ at the present time one
should have the constant $C$ to be fixed somewhere near $C = 3$ with an
accuracy of few percent. This is a fine-tuning, which does not sound very
attractive. However, it is important to realize that we are not talking about
an exponentially good precision; accuracy of few percent is good enough.

A similar result can be obtained even without changing the shape of the
effective potential. It is enough to assume that the field $\phi$ has a
nonminimal interaction with gravity of the form $-{\xi\over 2} R\phi^2$.  In
this case inflation becomes impossible for $\phi > {M_{\rm P}\over
\sqrt{8\pi\xi}}$
\cite{Maeda,BL95}. Thus in order to ensure that only closed inflationary
universes can be produced during the process of quantum creation of the
universe in the theory ${m^2\over 2} \phi^2$ it is enough to assume that
${M_{\rm P}\over \sqrt{8\pi\xi}} < 3M_{\rm P}$. This gives a condition $\xi >
{1\over 72\pi} \sim
4\times10^{-4}$.

To make sure that this mechanism of a closed universe creation is viable one
should check that the universe produced that way is sufficiently homogeneous.
Even though one may expect that this is guaranteed by the large absolute value
of
the gravitational action, one should check that the asymmetry of the
universe at the moment of its creation does not exceed an extremely small value
$\sim 10^{-5}$, since otherwise our mechanism will produce   anisotropy
of the microwave background radiation exceeding its experimentally established
value ${\Delta T\over T} \sim 10^{-5}$.

Calculation of probability of creation of a closed universe is a very
controversial subject,   and determination of its quantum state and of its
possible asymmetry  is even more complicated. However, one can  make a simple
estimate and show that the absolute value of the action $|S| = {3M^4_{\rm
P}\over 16V(\phi)} $ on the tunneling trajectory describing the universe
formation will change by $\Delta |S| \sim O(1)$ if one adds perturbation of the
standard amplitude $\delta \phi \sim {H\over 2\pi}$ to the homogeneous
background of the scalar field $\phi$. Trajectories with $\Delta |S| < 1$ are
not strongly suppressed as compared with the original tunneling trajectory.
Therefore tunneling into configurations with
$\delta \phi \sim {H\over 2\pi}$ can be possible, whereas we expect that
tunneling with creation of the universes with $\delta \phi \gg  {H\over 2\pi}$
should be exponentially suppressed as compared with the tunneling with creation
of the universes with $\delta \phi \sim {H\over 2\pi}$.

This result suggests
that expected deviations of the scalar field from homogeneity at the moment of
the universe
creation have the usual quantum mechanical amplitude ${H\over 2\pi}$ which is
responsible for
galaxy formation and anisotropy of the microwave background radiation in the
standard versions of inflation in a flat universe. In addition to this, there
will be certain irregularities of the shape of the original bubble, since its
size $\sim H^{-1}$ at the tunneling
is defined with an accuracy  $\sim M^{-1}_{\rm P}$. The resulting anisotropy
$\sim
{H\over
M_{\rm P}}$ is similar to the amplitude of gravitational waves produced during
inflation. In other words, both scalar and gravitational perturbations induced
at the moment of the universe creation are expected to  be
{\it of  the same magnitude}  as if the universe were inflating for an
indefinitely long time. Therefore tunneling may play the same role as inflation
from the point of view of the homogeneity and isotropy problems
\cite{Lab,Omega}.

 This possibility is very intriguing.  Still, we do not want to insist that
our conclusions are unambiguous. For example,  one could
argue that it is much
more natural for the universe to be created with the density very close to the
Planck density. However, the effective potential (\ref{1}) at the Planck
density is extremely steep.  Therefore such a universe   will not typically
enter the
inflationary regime, and will recollapse within a very short time.  It could
survive long enough for
inflation to occur  only if it was extremely large and relatively homogeneous
from the very beginning. If the probability of  creation of such a large
universe is
smaller than the probability of a direct creation of a homogeneous closed
inflationary universe which we studied above, all our conclusions will remain
intact.  Some preliminary estimates of the probability of creation of a large
universe which subsequently enters the stage of inflation can be found in
\cite{DA}; however, this issue  requires a more detailed  investigation.

Leaving apart this cautious note, our main conclusion is that it may be
possible to make inflation short and the universe
closed and homogeneous. The
remaining problem is to understand why our universe does not have $\Omega =
100$?
But in fact it is very easy to   answer this question: Value of $\Omega$
changes in a closed universe while it expands. It spends only a small fraction
of its lifetime in a state with $\Omega \gg 1$. About a half of its lifetime
before the closed universe
becomes collapsing it has $\Omega$ only slightly greater than 1. Therefore a
considerable fraction of all observers who may populate a closed universe
should live there at the time when $\Omega$ is not much greater than $1$. It is
as
simple as that. The situation with an open universe is much more complicated,
since an open universe spends only a finite amount of time at the beginning of
its evolution in a state with $\Omega \sim 1$, and then $\Omega$ decreases and
stays for indefinitely long time in a state with  $\Omega \ll 1$ ($\Omega \to
0$ for $t \to \infty$).

\section{\label{Onefield} Open universe in the models with  one scalar
field}

As we have already mentioned in the Introduction, the way to obtain an open
homogeneous inflationary universe is to use the mechanism outlined in
\cite{CL}--\cite{BGT}.  An open universe corresponds to an interior  of a
single bubble
appearing in the decaying false vacuum. This picture can be consistent with
observations only if the probability of tunneling with the bubble formation is
sufficiently small, so that the bubbles do not collide until the typical size
of the bubble becomes greater than the size of the observable part of our
universe. The corresponding constraints  are very easy to satisfy in the
theories with small coupling constants, where the tunneling rate is very small
\cite{Gott}. However, if one modifies the theory in such a way that the
probability of the bubble formation at some moment becomes so large that the
phase transition completes in the whole universe (see e.g. \cite{Occh}), then
there will be a danger that an observer inside the bubble will see
inhomogeneities created by other bubbles. Therefore will not study here
theories of such type.

It is not very easy to find a reasonable model which will lead to
tunneling and inflation by about $e^{60}$ times after the bubble formation. The
simplest idea   \cite{Bucher} is to realize this scenario in  the chaotic
inflation model with the effective potential
\begin{equation}\label{o1}
V(\phi) = {m^2\over 2} \phi^2 - {\delta\over 3} \phi^3 + {\lambda\over 4}\phi^4
\ .
\end{equation}
In order to obtain an open inflationary
universe in this model it is necessary to adjust   parameters  in such a way as
to ensure that the
tunneling creates bubbles containing the field    $\phi \sim 3 M_{\rm P}$. In
such a case the
interior of the bubble after its formation inflates by about $e^{60}$ times,
and $\Omega$ at the present time may become equal to, say, $0.3$. This
requires a fine tuning of the effective potential. If, for example, tunneling
occurs not to $\phi \sim 3 M_{\rm P}$ but to $\phi \sim 3.1 M_{\rm P}$ then the
universe will become practically flat.   It is worth noting, however, that
the required fine tuning is about the same order as for the closed universe
model described in Section \ref{Closed}, i.e.\ few percent.

Fine tuning is not the main difficulty of this  model. The tunneling
should occur to the part of the potential which   almost does not change its
slope during inflation at smaller $\phi$, since otherwise one does not   obtain
scale-invariant density
perturbations. One of the necessary conditions is that the barrier should be
very narrow. Indeed, if $V'' \ll H^2$ at the barrier, then the tunneling occurs
to its  top, as in the Hawking-Moss case \cite{HM}. Original interpretation of
this possibility was rather obscure; understanding came after the development
of stochastic approach to inflation. What happens is that the inflaton field
due to long-wave quantum fluctuations experiences Brownian motion. Occasionally
this field may jump from the local minimum of the effective potential to its
local maximum, and then slowly roll down from the maximum to the global minimum
\cite{Star,Gonch,Lab}.   If this
happens, one obtains unacceptably large density perturbations
${\delta\rho\over \rho} \sim {H^2\over \dot\phi} > 1$ on the large scale, since
$\dot\phi = 0$ at
the local maximum of $V(\phi)$.

Unfortunately, this is exactly the case for the model (\ref{o1}) \cite{Bucher}.
Indeed, the local minimum of the effective potential in this model appears at
\begin{equation}\label{o2}
\phi = {\delta \over 2 \lambda} + \gamma \ ,
\end{equation}
where
\begin{equation}\label{o3}
 \gamma = \sqrt{  {\delta^2 \over 4 \lambda^2} - {m^2\over \lambda} } \ .
\end{equation}
The local  minimum of the effective potential appears for ${\delta > 2
\sqrt\lambda\, m }$, and it becomes unacceptably deep (deeper than the minimum
at $\phi = 0$) for  ${\delta > {3 \sqrt\lambda\over \sqrt 2}\, m }$.  Thus in
the whole region of interest one can use a simple estimate ${\delta \sim  2
\sqrt\lambda\, m }$ and represent $\gamma$ in the form $\beta{m\over 2
\sqrt{2\lambda}}$, with  $\beta < {1}$.  The local maximum of the potential
appears at $\phi = {\delta \over 2 \lambda} - \gamma$. Tunneling should occur
to some point with $3 M_{\rm P} < \phi < {\delta \over 2 \lambda} - \gamma$,
which
implies that ${\delta \over  \lambda} > 6 M_{\rm P}$.

The best way to study tunneling in this theory is to introduce the field $\chi$
in such a way that $\chi = 0$ at the local minimum of $V(\phi)$:
\begin{equation}\label{o4}
\chi = -\phi +{\delta \over 2 \lambda} + \gamma \ .
\end{equation}
After simple algebra one can show that  if the local minimum is not very deep
($\beta \ll 1$), the effective potential (\ref{o1}) can be represented as
\begin{equation}\label{o5}
V(\chi) \sim {m^2\delta^2 \over 48\lambda^2} + { \sqrt 2\beta m^2 } \chi^2 -
{\delta\over 3} \chi^3 + {\lambda\over 4}\chi^4 \ .
\end{equation}
The Hubble constant in the local minimum  is given by
$H^2 \sim  { \pi \delta^2  m^2\over 18\lambda^2 M_{\rm P}^2} >  {2\pi}\, m^2$,
which
is much greater than the effective mass squared of the field $\chi$ for
$\beta \ll 1$, $m^2_\chi = {2\sqrt 2\beta m^2}$. In this regime tunneling
should occur to the local maximum of the effective potential, which should lead
to   disastrous consequences for the spectrum of density perturbations.

A possible way to overcome this problem is to consider the case ${\delta
\approx {3 \sqrt\lambda\over \sqrt 2}\, m }$\, ($\beta \approx 1$). Then the
two minima of the effective potential become nearly degenerate in energy, and
the Hubble constant becomes much smaller than $m_\chi$. (In this regime our
estimate for $H$, which was valid for $\beta \ll 1$,  should be improved.)
However, in such a situation we will have other problems. In this regime
tunneling occurs almost exactly to the minimum of the effective potential at
$\phi = 0$. Therefore it becomes difficult to have any inflation at all after
the tunneling, and the problem of fine tuning becomes especially severe.

Note that this problem is rather general. Its  origin  is in the condition that
in the   inflationary universe scenario the curvature of the effective
potential (the mass squared of the inflaton field)  $60$
e-foldings prior to the end of inflation always is much smaller than $H^2$.
Therefore one should bend the effective potential in a quite dramatic way in
order to create a local minimum at large $\phi$ and to make the curvature of
the effective potential in this minimum much  greater than $H^2$.  One can
avoid this problem  by introducing non-minimal kinetic terms
in the Lagrangian of  the inflaton field \cite{Bucher}, but this is just
another representation of the artificial bending of the effective potential.
Of course, it may happen that the bending of the effective potential  can
appear in a natural way in the theories based on supergravity and superstrings.
The simplest idea is to multiply the effective potential (\ref{o1}) by the
factor of the type $\exp{\Bigl({\phi\over CM_{\rm P}}\Bigr) } $ or
$\exp{\Bigl({\phi\over CM_{\rm P}}\Bigr)^2} $, like in Eq.\ (\ref{1}). At small
$\phi$ these factors will not influence the shape of the effective potential,
but at large $\phi$ they will make it   very curved.  This is
exactly what we need to avoid the Hawking-Moss tunneling.
Still the necessity to make all these tricks with bending   the potential and
making it very curved at some fine-tuned value of the scalar field $\phi$ do
not make the models of this type particularly attractive.

Therefore in the next sections we will make an attempt
to find some simple models where an open inflationary universe can appear in a
more
natural way. However, before doing so we will consider one more problem which
should be addressed in all versions of the open inflationary universe scenario.

\section{\label{Bubbles} Tunneling probability and spherical symmetry}

In the previous section we have assumed that the bubbles are exactly
spherically symmetric (or, to be more accurate, $O(3,1)$-symmetric \cite{CL}).
Meanwhile in realistic situations this condition may be violated for several
reasons. First of all, the bubble may be formed not quite symmetric. Then  its
shape may change even further due to growth of its initial  inhomogeneities and
due to quantum fluctuations which appear during the bubble wall expansion. As
we will see, this may cause a lot of problems if one wishes to maintain the
degree of anisotropy of the microwave background radiation inside the bubble at
the level of $10^{-5}$.

First of all, let us consider the issue of symmetry of a bubble at the moment
of its formation. For simplicity we will investigate the models where tunneling
can be described in the thin wall approximation. We will neglect gravitational
effects, which is possible as far as the initial radius $r$ of the bubble is
much smaller than $H^{-1}$. In this approximation (which works rather well for
the models to be discussed)
euclidean action of the $O(4)$-symmetric instanton describing bubble formation
is given by
\begin{equation}\label{o6}
S  = - {\epsilon\over 2} \pi^2 r^4 + 2\pi^2 r^3 s \  .
\end{equation}
Here $r$ is the radius of the bubble at the moment of its formation, $\epsilon$
is the difference of $V(\phi)$ between the false vacuum $\phi_{\rm initial}$
and the true vacuum $\phi_{\rm final}$, and $s$ is the surface tension,
\begin{equation}\label{o7}
s = \, \int_{\phi_{\rm initial}}^{\phi_{\rm final}} \sqrt{ 2(V(\phi) -
V(\phi_{\rm
final}))}\,  d\phi \  .
\end{equation}
The radius of the bubble can be obtained from the extremum of  (\ref{o6})
with respect to $r$:
\begin{equation}\label{o8}
r = {3s\over \epsilon } \ .
\end{equation}
Let us check how the action $S $ will change if one consider a bubble of a
radius $r + \Delta r$. Since the first derivative of $S $ at its extremum
vanishes, the change will be determined by its second derivative,
\begin{equation}\label{09}
\Delta S = {1\over 2} S'' (\Delta r)^2 = 9\pi^2\, {s^2\over \epsilon}\, (\Delta
r)^2 \ .
\end{equation}
 Now we should remember that all trajectories which have an action different
from the action at extremum by no more than  $1$ are quite legitimate. Thus the
typical deviation of the radius of the bubble from its classical value
(\ref{o8}) can be estimated from the condition $\Delta S \sim 1$, which gives
\begin{equation}\label{o10}
|\Delta r| \sim {\sqrt\epsilon\over 3\pi \,s} \ .
\end{equation}
Note, that even though we considered spherically symmetric perturbations, our
estimate  is based on corrections proportional to $(\delta r)^2$, and
therefore it should remain valid for perturbations which have an amplitude
$\Delta r$, but change their sign in different parts of the bubble surface.
Thus, eq. (\ref{o10}) gives an estimate of a typical degree of asymmetry of the
bubble at the moment of its creation:
\begin{equation}\label{o11}
A(r) \equiv  {|\Delta r| \over r} \sim {\epsilon\sqrt\epsilon\over 3\pi \,s^2}
\ .
\end{equation}
This simple estimate exactly coincides with the corresponding result obtained
by Garriga and Vilenkin \cite{VilGarr} in their study of quantum fluctuations
of bubble walls. It was shown in \cite{VilGarr} that when an   empty  bubble
begins
expanding, the typical deviation $\Delta r$ remains constant. Therefore the
asymmetry given by the ratio ${|\Delta r| \over r}$ gradually vanishes. This is
a pretty general result: Waves produced by a brick falling to a pond do not
have the shape of a brick, but gradually become circles.

However, in our case the situation is somewhat more complicated. The wavefront
produced by a brick in inflationary background preserves the shape of the brick
if its size  is much greater than $H^{-1}$. Indeed, the wavefront moves   with
the speed approaching the speed of light, whereas the distance between
different parts of a region with initial size greater than $H^{-1}$ grows with
a much greater (and ever increasing) speed. This means that inflation
stretches the wavefront without changing its shape on scale much greater than
$H^{-1}$. Therefore during inflation which
continues inside the bubble the symmetrization of its shape   occurs only in
the very beginning, until the radius of the bubble approaches $H^{-1}$. At this
first stage expansion of the bubble occurs mainly due to the motion of the
walls rather than due to inflationary stretching of the universe, and our
estimate of the bubble wall asymmetry as well as the results obtained by
Garriga and Vilenkin for the empty bubble remain valid. At the moment when the
radius of the bubble becomes equal to $H^{-1}$ its asymmetry
becomes
\begin{equation}\label{o12}
A(H^{-1}) \sim {|\Delta r| H} \sim {\sqrt\epsilon H\over 3\pi \,s} \ ,
\end{equation}
and the subsequent expansion of the bubble does not change this value very
much. Note that the Hubble constant here is determined by the vacuum energy
{\it
after} the tunneling, which may differ from the initial energy density
$\epsilon$.

The deviation of the shape of the
bubble from spherical symmetry implies that the beginning of the second stage
of inflation inside the bubble will be not exactly synchronous, with the delay
time $\Delta t \sim \Delta r$. This, as usual,  may lead to adiabatic density
perturbations on the
horizon scale of the order of $H\Delta t$, which coincides with the bubble
asymmetry $A$ after its size becomes greater than $H^{-1}$, see Eq.\
(\ref{o12}).

To estimate this contribution to density perturbations, let us consider again
the
simplest model with the effective potential (\ref{o1}). Now we will consider it
in the limit  $\beta - 1 \ll 1$ which implies that the two minima have almost
the same depth, which is necessary for  validity of the thin wall
approximation. In this case    $2\delta^2 = 9 M^2\lambda$, and the effective
potential (\ref{o1}) looks approximately like ${\lambda \over 4} \phi^2 (\phi
- \phi_0)^2$, where $\phi_0 = {2\delta\over 3\lambda} = \sqrt {2\over \lambda}
{M}$ is the position of the local minimum of the effective
potential.  The surface tension in this model is given by $s =
\sqrt{\lambda\over 2} {\phi_0^3\over 6} = { M^3\over 3\lambda}$  \cite{Tunn}.
We will also introduce a phenomenological parameter $\mu$, such that $\mu
{M^4\over 16\lambda} = \epsilon$. The smallness of this parameter controls
applicability of the thin-wall approximation, since the value of the effective
potential near the top of the potential barrier   at $\phi = \phi_0/2$ is given
by $M^4\over 16\lambda$.   Then our estimate of density
perturbations associated with the bubble wall (\ref{o12})    gives
\begin{equation}\label{o14}
\left. {\delta\rho\over \rho}\right|_{\rm bubble}  \sim A(H^{-1}) \sim
{\sqrt{\mu\lambda}
H\over  4 \pi
M}   \ .
\end{equation}
Here $H$ is the value of the Hubble constant at the beginning of inflation
inside the bubble.

In order to have  $\left. {\delta\rho\over \rho}\right|_{\rm bubble} {\
\lower-1.2pt\vbox{\hbox{\rlap{$<$}\lower5pt\vbox{\hbox{$\sim$}}}}\ } 5 \times
10^{-5}$ (the number $5 \times 10^{-5}$ corresponds to the amplitude of density
perturbations in  the COBE normalization) one
should have
\begin{equation}\label{o17}
\left. {\delta\rho\over \rho}\right|_{\rm bubble}  \sim \, {\sqrt{\mu\lambda}
H\over  4 \pi
M}\,  {\ \lower-1.2pt\vbox{\hbox{\rlap{$<$}\lower5pt\vbox{\hbox{$\sim$}}}}\ }\,
 5 \times 10^{-5} \ .
\end{equation}
For $H\ll M$ perturbations produced by the bubble walls   may   be sufficiently
small  even if the coupling constants are relatively large and the bubbles at
the moment of their formation are very inhomogeneous.

There is a long way from our simple estimates  to the full theory of
anisotropies of
cosmic microwave background induced by fluctuations of the domain wall. In
particular, the significance of this effect will clearly depend on the value of
$\Omega$.
 The constraint (\ref{o17}) may appear  only if one can ``see'' the
scale at which the bubble walls have imprinted their fluctuations. If inflation
is long enough,
this scale becomes
exponentially large, we do not see the fluctuations due to bubble walls, but
then we return to the standard
inflationary scenario of a flat inflationary universe. However, for $\Omega \ll
1$ inflation is short, and it does not preclude us from seeing perturbations in
a vicinity of the bubble walls \cite{Open}.\footnote{One should  distinguish
between
the
infinite size of an open universe and the finite  distance from us to the
bubble walls along the light cone.}
In such a case one should   take
the constraint  (\ref{o17}) very seriously.

In the open universe the form of the spectrum of CMBR temperature
fluctuations can be substantially different from
the form of the spectrum of density fluctuations because of the integral
Sachs-Wolfe effect \cite{Open} (see, in particular, the paper by Lyth and
Woszczyna and references therein). In addition to this, the perturbations
discussed above occur on  a super-curvature scale. Therefore, they provide a
natural source for the  Grishchuk-Zeldovich effect (according to Lyth and
Woszczyna
 \cite{Open} only the modes which are not in conformal vacuum can be
 responsible for this), while the density fluctuations produced during the
 secondary inflation are not likely to contain super-curvature modes.

One can show that in the theories of   one
scalar field  similar to the model discussed in the previous section (if this
model would work)  the condition (\ref{o17}) is almost automatically satisfied.
Meanwhile in the
models of two different scalar fields, which we are going to discuss now, this
condition may lead
to additional restrictions on the parameters of the models.

\section{\label{Simplest} The simplest model of a (quasi)open inflationary
universe}

As we have seen in Sect. \ref{Onefield}, it is rather difficult to obtain an
open universe in the models of one scalar field with simple potentials, such
as, e.g., ${m^2\over 2} \phi^2 - {\delta\over 3} \phi^3 + {\lambda\over
4}\phi^4$.  In this section we will explore an extremely  simple model of two
scalar fields, where the universe after inflation becomes open (or quasiopen,
see below) in a very natural way \cite{Omega}.

Consider a model of
two noninteracting scalar fields, $\phi$ and $\sigma$, with the effective
potential
\begin{equation}\label{3}
V(\phi, \sigma) = {m^2\over 2}\phi^2 + V(\sigma) \ .
\end{equation}
Here $\phi$ is a weakly interacting inflaton field, and $\sigma$, for example,
can be the field responsible for the symmetry breaking in GUTs. We will assume
that $V(\sigma)$ has a local minimum at $\sigma = 0$, and a global minimum at
$\sigma_0 \not = 0$, just as in the old inflationary
theory. For definiteness, we will assume that this potential is given by
${M^2\over 2} \sigma^2 -
{\alpha M } \sigma^3 + {\lambda\over 4}\sigma^4 + V(0)$, with $V(0) \sim
{M^4\over 4 \lambda}$, but it is not essential;
no fine tuning of the shape of this potential will be required.

Note that so far we did not make any unreasonable complications to the standard
chaotic inflation scenario; at large $\phi$ inflation is driven
by the field $\phi$, and the GUT potential is necessary in the theory anyway.
In order to obtain density perturbations of the necessary amplitude the mass
$m$ of the scalar field $\phi$ should be of the order of $10^{-6} M_{\rm P}
\sim
10^{13}$ GeV \cite{MyBook}.

Inflation begins at $V(\phi, \sigma) \sim M_{\rm P}^4$. At this stage
fluctuations of
both fields are very strong, and the universe enters the stage of
self-reproduction, which finishes for the field $\phi$ only when it becomes
smaller than $M_{\rm P} \sqrt{M_{\rm P}\over m}$ and the energy density drops
down to $m
M_{\rm P}^3  \sim 10^{-6} M_{\rm P}^4$ \cite{MyBook}. Quantum fluctuations of
the field
$\sigma$ in some parts of the universe put it directly to the absolute minimum
of
$V(\sigma)$, but in some other parts the scalar field $\sigma$ appears in the
local minimum of $V(\sigma)$ at $\sigma  = 0$. We will follow evolution of such
domains. Since the energy density in such
domains will be greater, their volume will  grow  with a greater speed, and
therefore they will be especially important for us.

One may worry that all
domains with $\sigma = 0$
will tunnel to the minimum of $V(\sigma)$ at the stage when the field $\phi$
was very large and quantum fluctuations of the both fields were large too.
This may happen if the Hubble constant induced by the scalar field $\phi$ is
much greater than the curvature of the potential $V(\sigma)$:
\begin{equation}\label{s1}
{m\phi\over M_{\rm P}} {\
\lower-1.2pt\vbox{\hbox{\rlap{$>$}\lower5pt\vbox{\hbox{$\sim$}}}}\ } M \ .
\end{equation}

This decay can be easily suppressed if one introduces a
small interaction $g^2\phi^2\sigma^2$ between these two fields, which
stabilizes the state with $\sigma = 0$ at large $\phi$. Another possibility,
which we have already mentioned in Sect. \ref{Closed}, is to add a
nonminimal interaction with gravity of the form $-{\xi\over 2} R\phi^2$, which
makes inflation impossible for $\phi > {M_{\rm P}\over 8\phi\xi}$. In this case
the
condition (\ref{s1}) will never be satisfied.  However, there is a much simpler
answer to this worry. If the effective potential of the field $\phi$ is so
large that the field $\phi$ can easily jump to the true minimum of $V(\sigma)$,
then the universe becomes divided into infinitely many domains with all
possible values of $\sigma$ distributed in the following way
\cite{Star,MyBook}:
\begin{equation}\label{s2}
{P(\sigma= 0)\over P(\sigma = \sigma_0)} \sim \exp\left({3M^4_{\rm P}\over 8
V(\phi,0)} - {3M^4_{\rm P}\over 8V(\phi,\sigma)}\right) = \exp\left({3M^4_{\rm
P}\over 4(m^2\phi^2 + 2V(0))} - {3M^4_{\rm P}\over 4 m^2\phi^2}\right)\ .
\end{equation}
One can easily check that at the moment when the field $\phi$ decreases to ${M
M_{\rm P}\over m}$ and  the condition (\ref{s1}) becomes violated, we will
have
\begin{equation}\label{s3}
{P(0)\over P(\sigma_0)}  \sim \exp\left(-{C\over \lambda}\right) \ ,
\end{equation}
where $C$ is some constant, $C = O(1)$. After this moment the probability of
the false vacuum decay typically becomes much smaller. Thus the fraction of
space which survives in the false vacuum state $\sigma = 0$ until this time
typically is very small, but finite (and calculable). It is important,   that
these rare domains with $\sigma = 0$ eventually will dominate the volume of the
universe since if the probability of the false vacuum decay is small enough,
the volume of the domains in the false vacuum will continue growing
exponentially without end.

The main idea of our scenario can be explained as follows. Because the fields
$\sigma$ and
$\phi$ do not interact with each other, and the dependence of the probability
of tunneling on the vacuum energy at the GUT scale is negligibly small
\cite{CL}, tunneling to the minimum of $V(\sigma)$ may occur with approximately
equal
probability at all sufficiently small values of the field $\phi$ (see, however,
below). The
parameters of the bubbles of the field $\sigma$ are determined by the mass
scale $M$ corresponding to the effective potential $V(\sigma)$. This mass scale
in
our model is much greater than $m$. Thus the duration of tunneling in the
Euclidean ``time'' is much smaller than $m^{-1}$. Therefore the field $\phi$
practically does not change its value during the tunneling.  If
the probability of decay at a given $\phi$ is small enough, then it does not
destroy the whole vacuum state $\sigma = 0$ \cite{GW}; the bubbles of the new
phase are produced all the way when    the field $\phi$ rolls down to $\phi =
0$. In this process  the universe  becomes filled with
(nonoverlapping) bubbles immersed in the false vacuum state with $\sigma = 0$.
Interior of each of these bubbles   represents an open universe. However, these
bubbles   contain {\it different} values of the field $\phi$, depending on the
value of this field at the  moment when the bubble formation occurred. If the
field $\phi$ inside a bubble is smaller than $3 M_{\rm P}$, then the universe
inside
this bubble will have a vanishingly small $\Omega$, at the age $10^{10}$ years
after the end of inflation it will be practically empty, and life of our type
would not exist there.  If the field $\phi$ is much greater than $3 M_{\rm P}$,
the
universe inside the bubble will be almost exactly flat, $\Omega = 1$, as in the
simplest version of the chaotic inflation scenario. It is important, however,
that {\it in  an eternally existing self-reproducing universe there will be
infinitely many universes containing any particular value of $\Omega$, from
$\Omega = 0$ to $\Omega = 1$}, and one does not need any fine tuning of the
effective potential to obtain a universe with, say,  $0.2 <\Omega < 0.3$

Of course, one can argue that we did not solve the problem of fine tuning, we
just transformed it into the fact that only a very small percentage of all
universes will have  $0.2 <\Omega < 0.3$. However, first of all, we
achieved our goal in a very simple theory, which does not require any
artificial potential bending and nonminimal kinetic terms. Then, there may be
some reasons why it is preferable for us to live in a universe with a small
(but not vanishingly small) $\Omega$.

The simplest way to approach this problem is to find  how the probability
for the bubble productiondepends on $\phi$. As we already pointed out, for
small $\phi$ this dependence is not very strong. On the other hand, at large
$\phi$ the probability rapidly grows and  becomes quite large at $\phi > {M
M_{\rm P}\over m}$. This may suggest that the bubble production typically
occurs at
$\phi > {M M_{\rm P}\over m}$, and then for ${M\over  m} \gg 3$ we typically
obtain
flat universes, $\Omega = 1$. This is another manifestation of the problem of
premature decay of the state $\sigma = 0$ which we discussed above. Moreover,
even if the probability to produce the universes with different $\phi$ were
entirely $\phi$-independent, one could argue that the main volume of the
habitable parts of the universe is contained in the bubbles with $\Omega = 1$,
since  the interior of each such bubble  inflated longer. Again, there exist
several ways of resolving this problem: involving coupling $g^2\phi^2\sigma^2$,
which stabilizes the state $\sigma = 0$ at large $\phi$, or adding nonminimal
interaction with gravity of the form $-{\xi\over 2} R\phi^2$, which makes
inflation impossible for $\phi > {M_{\rm P}\over \sqrt{8\pi\xi}}$. In either
way one can
easily suppress production of the universes with   $\Omega = 1$. Then the
maximum of probability will correspond to some value $\Omega < 1$, which can be
made equal to any given number from $1$ to $0$ by changing the parameters $g^2$
and $\xi$. \footnote{Thus we disagree with the statement  made in \cite{Occh}
that this model
typically predicts empty universes.}

However, calculation of probabilities in the context of the theory of a
self-reproducing universe is a very ambiguous process. For example, we may
formulate the problem in a different way. Consider a domain of the false vacuum
with $\sigma = 0$ and $\phi = \phi_1$. After some evolution it
produces one or many bubbles with $\sigma = \sigma_0$ and the field $\phi$
which after some time becomes equal to $\phi_2$. One may argue that the most
efficient way this process may go is the way which in the end produces the
greater volume. Indeed, for the inhabitants of a bubble it does not matter how
much time did it take for this process to occur. The total number of
observers produced by this process will depend on the total volume of the
universe at the hypersurface of a given density, i.e. on the hypersurface of a
given $\phi$. If the domain instantaneously  tunnels to the state $\sigma_0$
and $\phi_1$, and then the field $\phi$ in this domain slowly rolls from
$\phi_1$ to $\phi_2$, then the volume of this domain grows $\exp
\Bigl({2\pi\over M_{\rm P}^2} (\phi_1^2 -\phi_2^2)\Bigr)$ times \cite{MyBook}.
Meanwhile, if the tunneling takes a long time, then the field $\phi$ rolls down
extremely slowly being in the false vacuum state with $\sigma = 0$. In this
state the universe expands much faster than in the state with $\sigma =
\sigma_0$. Since it expands much faster, and it takes the field much longer to
roll from $\phi_1$ to $\phi_2$, the trajectories of this kind bring us much
greater volume. This may serve as an argument that most of the volume is
produced by the bubbles created at a very small $\phi$, which leads to the
universes with very small $\Omega$.

One may use another set of considerations, studying all trajectories beginning
at $\phi_1, t_1$ and ending at $\phi_2, t_2$. This will bring us  another
answer, or, to be more precise, another set of answers, which will depend on
the choice of the time parametrization \cite{LLM}.  A very interesting approach
was recently proposed by Vilenkin, who suggested to introduce a particular
cutoff procedure which (almost) completely eliminates dependence of the final
answer on the time parametrization \cite{VilNew}. A more radical possibility
would be to integrate over all time parametrizations. This task is very
complicated, but it   would completely eliminate dependence of the final answer
on the time parametrization \cite{OPEN}.

There is a very deep  reason why the calculation of the probability to obtain a
universe with a given $\Omega$ is so ambiguous. For those who will live inside
a bubble there will be no way to say at which stage (at which
time from the point of view of an external observer) this bubble was produced.
Therefore one should compare {\it all} of these bubbles produced at all
possible times.  The self-reproducing universe should exist for indefinitely
long time, and therefore   it should contain  infinitely many bubbles with all
possible values of $\Omega$. Comparing infinities is a very ambiguous task,
which gives results depending on the procedure of comparison. For example, one
can consider an infinitely large box of apples and an infinitely large box of
oranges. One may pick up one apple and one orange, then one apple and one
orange, over and over again, and conclude that there is an equal number of
apples and oranges. However, one may also pick up one apple and
two oranges, and then one apple and two oranges again, and conclude that there
is twice as many oranges as apples. The same situation happens when one tries
to compare the number of  bubbles with different values of $\Omega$. If we
would know how to
solve  the problem of measure in quantum cosmology, perhaps we would be able to
obtain
something similar to an open universe in the trivial $\lambda\phi^4$ theory
without any first order phase transitions
\cite{OPEN}.  In the meantime, it is already encouraging that in our scenario
there are infinitely many inflationary universes with all possible value of
$\Omega < 1$. We can hardly live in the empty bubbles with $\Omega = 0$. As for
the choice between the bubbles with different nonvanishing values of $\Omega <
1$,  it is quite possible that eventually we will find out an unambiguous way
of predicting the most probable value of $\Omega$, and we are going to continue
our work in this direction. However,  it might also happen that this question
is as meaningless as the question whether it is more probable to be born as a
chinese rather than as an italian. It is quite conceivable that the only way to
find out in which of the bubbles do we   live   is to make observations.

Some words of caution are in order here. The bubbles produced in our simple
model
are not {\it exactly} open universes. Indeed, in the models discussed in
\cite{CL}--\cite{BGT} the time of reheating (and the temperature of the
universe after the reheating) was synchronized with the value of the scalar
field inside the bubble. In our case the situation is very similar, but not
exactly. Suppose that the Hubble constant induced by $V(0)$ is much
greater than the Hubble constant related to the energy density of the scalar
field $\phi$. Then the speed of rolling of the scalar field $\phi$ sharply
increases inside the bubble. Thus, in our case the field $\sigma$ synchronizes
the motion of the field $\phi$, and then the hypersurface of a constant field
$\phi$ determines the hypersurface of a constant temperature. In the models
where the rolling of the field $\phi$ can occur only inside the bubble (we will
discuss such a model shortly)  the  synchronization is precise, and everything
goes as in the models of refs. \cite{CL}--\cite{BGT}. However, in our simple
model the scalar field $\phi$ moves down outside the bubble as well, even
though it does it very slowly. Thus,  synchronization of   motion of the
fields $\sigma$ and $\phi$  is not precise; hypersurface of a constant $\sigma$
ceases to be a hypersurface of a constant density. For example, suppose that
the field $\phi$ has taken some value $\phi_0$ near the bubble wall when the
bubble was just formed. Then the bubble expands, and during this time the field
$\phi$ outside the wall  decreases, as $\exp \Bigl(-{m^2t\over 3 H_1}\Bigr)$,
where $H_1 \approx  H(\phi = \sigma = 0)$ is the Hubble constant at the first
stage of inflation, $H_1 \approx \sqrt{8\pi V(0)\over 3 M_{\rm P}^2}$
\cite{MyBook}. At the moment
when the bubble expands $e^{60}$ times, the field $\phi$ in the region just
reached by  the bubble wall decreases to  $\phi_o\exp \Bigl(-{20 m^2\over
H^2_1}\Bigr)$ from its original value $\phi_0$. The universe inside the bubble
is a homogeneous open universe only if this change is negligibly small. This
may not be a real problem. Indeed,  let us assume that $V(0) ={\tilde M}^4$,
where ${\tilde M} =
10^{17}$ GeV. (Typically the energy density scale $\tilde M$ is related to the
particle mass as follows: ${\tilde M} \sim \lambda^{-1/4} M$.) In this case
$H_1 = 1.7 \times 10^{15}$ GeV, and for $m =
10^{13}$ GeV one obtains ${20 m^2\over   H_1^2} \sim 10^{-4}$. In such a case
a typical degree of distortion of the picture of a homogeneous open universe is
very small.

Still this issue requires careful investigation. When the bubble wall continues
expanding even further, the scalar field outside of it eventually drops down to
zero. Then there will be no new matter created near the wall.  Instead of
infinitely large homogeneous open universes we are obtaining   spherically
symmetric islands of a size much greater than the size of the observable part
of our universe. We do not know whether this unusual picture is an advantage or
a
disadvantage of our model. Is it possible to consider different parts of the
same
exponentially large island as domains of different ``effective'' $\Omega$? Can
we attribute some part of the dipole anisotropy of the microwave background
radiation to the possibility that we live somewhere outside of the center of
such island? In any
case, as we already mentioned, in the limit $m^2 \ll H_1^2$
we do not expect that the small deviations of the geometry of space inside the
bubble from the geometry of an open universe can do much harm to our model.

Our model admits many generalizations, and details of the scenario which we
just discussed depend on the values of parameters. Let us forget for a moment
about all complicated processes which occur  when the field $\phi$ is rolling
down to $\phi = 0$, since this part of the picture depends on the validity of
our ideas about initial conditions. For example, there may be no
self-reproduction of inflationary domains with large $\phi$ if one considers an
effective
potential of the field $\phi$ which is very curved at large $\phi$, as in eq.
(\ref{1}). However, there will be self-reproduction of the universe in a state
$\phi = \sigma = 0$, as in the old inflation scenario. Then the main portion of
the volume of the universe will be determined by the processes which occur when
   the fields    $\phi$  and $\sigma$    stay  at the local minimum of the
effective potential, $\phi = \sigma = 0$.   For definiteness we will assume
here that $V(0) = {\tilde M}^4$, where ${\tilde M}$ is   the   stringy scale,
${\tilde M} \sim 10^{17} -
10^{18}$ GeV. Then the Hubble constant $H_1 = \sqrt{8\pi V(0)\over 3M^2_{\rm
P}} \sim
\sqrt{8\pi \over 3} {{\tilde M}^2\over M_{\rm P}}$ created by the energy
density
$V(0)$ is
much greater than $m \sim 10^{13}$ GeV. In such a case the scalar field $\phi$
will not stay exactly at $\phi = 0$. It will  be  relatively homogeneous on the
horizon scale $H_1^{-1}$, but otherwise it will  be chaotically distributed
with
the dispersion $\langle\phi^2\rangle = {3H^4\over 8\pi^2m^2}$  \cite{MyBook}.
This means that the field $\phi$ inside each of the bubbles produced by the
decay of the false vacuum can take any value $\phi$ with the probability
\begin{equation}\label{4}
P \sim \exp\left(-{\phi^2\over 2 \langle\phi^2\rangle}\right) \sim
\exp\left(-{3m^2 \phi^2M_{\rm P}^4\over 16 {\tilde M}^8}\right) \ .
\end{equation}
One can check that for ${\tilde M} \sim 4.3\times10^{17}$ GeV the typical value
of the
field $\phi$ inside the bubbles will be $\sim 3\times 10^{19}$ GeV. Thus, for
${\tilde M} > 4.3\times10^{17}$ GeV most of the universes produced during the
vacuum
decay will be flat, for ${\tilde M} < 4.3\times10^{17}$ GeV most of them will
be open.
It is interesting that in this version of our model the percentage of open
universes is determined by the stringy scale (or by the GUT scale). However,
since the process of bubble production in this scenario   goes without   end,
the total  number of universes with any particular value of  $\Omega < 1$ will
be infinitely large   for any value of ${\tilde M}$.   Thus this   model shows
us is the
simplest way to resurrect some of the ideas of the old inflationary theory with
the help of chaotic inflation, and simultaneously to obtain  inflationary
universe with $\Omega < 1$.

Note that this version of our model will not suffer for the problem of
incomplete synchronization. Indeed, the average value of the field $\phi$ in
the false vacuum outside the bubble will remain constant until the bubble
triggers its decrease.
However, this model, just as its previous version, may suffer from another
problem. The Hubble constant $H_1$ before the tunneling in this model was much
greater
than the Hubble constant $H_2$ at the beginning of the second stage of
inflation. Therefore the fluctuations of the
scalar field before the tunneling were very large, $\delta \phi \sim {H_1\over
2 \pi}$, much greater than the
fluctuations generated after the tunneling,  $\delta \phi \sim {H_2\over 2
\pi}$. This may lead to very large
density perturbations on the scale comparable to the size of the bubble. For
the models with $\Omega = 1$ this effect would not cause any problems since
such
perturbations would be far away over the present particle horizon, but for
small $\Omega$  this
may lead to unacceptable anisotropy of the microwave background radiation.

Fortunately, this may not be a real difficulty. A possible solution is very
similar to the bubble symmetrization described in the previous section.

Indeed, let us consider more carefully how the long wave perturbations produced
outside the bubble may penetrate into it. At the moment when the bubble is
formed, it has a   size (\ref{o3}), which is  smaller than $H_1^{-1}$
\cite{CL}. Then the bubble walls begin moving with the speed gradually
approaching the speed of light. At this stage the comoving size of the bubble
(from the point of view of the original coordinate system in the false vacuum)
grows like
\begin{equation}\label{n1}
r(t) = \int_{0}^{t}{dt e^{-H_1 t}} = H_1^{-1} (1 - e^{-H_1 t}) \ .
\end{equation}
During this time the fluctuations of the scalar field $\phi$ of the amplitude
${H_1\over 2\pi}$ and of the wavelength $H_1^{-1}$, which previously were
outside the bubble, gradually become covered by it. When these perturbations
are outside the bubble, inflation with the Hubble constant $H_1$ prevents them
from oscillating and moving. However, once these perturbations penetrate inside
the bubble, their amplitude becomes decreasing \cite{MZ,SP}. Indeed, since the
wavelength of the perturbations is $\sim H_1^{-1} \ll H_2^{-1} \ll m^{-1}$,
these
perturbations move inside the bubbles as relativistic particles, their
wavelength grow  as $a(t)$, and their amplitude decreases just like an
amplitude of electromagnetic field, $\delta\phi \sim a^{-1}(t)$, where $a$ is
the scale factor of the universe inside a bubble \cite{MZ}. This process
continues until the wavelength of each perturbation reaches $H_2^{-1}$ (already
at the second stage of inflation). During this time the wavelength grows
${H_1\over H_2}$ times, and the amplitude decreases ${H_2\over H_1}$ times, to
become the standard amplitude of perturbations produced at the second stage of
inflation: $  {H_2\over H_1}\, {H_1\over 2\pi} = {H_2\over 2\pi}$.

In fact, one may  argue that this computation was too naive, and that these
perturbations should be neglected altogether. Typically we treat long wave
perturbations in inflationary universe like classical wave for the reason that
the waves with the wavelength much greater than the horizon can be interpreted
as states with extremely large occupation numbers \cite{MyBook}. However, when
the new  born perturbations (i.e. fluctuations which did not acquire an
exponentially large wavelength yet) enter  the bubble (i.e. under the horizon),
they effectively return to the realm of quantum fluctuations again. Then one
may argue that one should simply forget about the waves with the wavelengths
small enough to fit into the bubble, and consider perturbations created at the
second stage of inflation not as a result of stretching of these waves, but as
a new process of creation of perturbations of an amplitude ${H_2\over 2\pi}$.

One may worry   that perturbations which had wavelengths somewhat greater than
$H_1^{-1}$ at the moment of the bubble formation  cannot completely penetrate
into the bubble. If, for example, the field $\phi$ differs from some constant
by $+{H_1\over 2\pi}$ at the distance $H_1^{-1}$ to the left   of the bubble at
the moment of its formation, and by  $-{H_1\over 2\pi}$ at the distance
$H_1^{-1}$ to the  right of the bubble, then this difference remains frozen
independently of all processes inside the bubble. This may suggest that there
is some unavoidable asymmetry of the distribution of the field inside the
bubble. However, the field inside the bubble will not be distributed like a
straight line slowly rising from  $-{H_1\over 2\pi}$ to  $+{H_1\over 2\pi}$.
Inside
the bubble the field will be almost homogeneous; the inhomogeneity $\delta \phi
\sim -{H_1\over 2\pi}$ will be concentrated only in a small vicinity near the
bubble wall.

Finally we should verify that this scenario  leads to bubbles which are
symmetric enough, see eq. (\ref{o17}). Fortunately, here we do not have any
problems. One can easily check that for our model with $m \sim 10^{13}$ GeV and
$\tilde M \sim \lambda^{-1/4} M > 10^{17} GeV$ the condition (\ref{o17}) can be
  satisfied even for not very small values of the coupling constant $\lambda$.

The arguments presented above should be confirmed by a more detailed
investigation of the vacuum structure inside the expanding bubble in our
scenario. If, as we hope,  the result of the investigation will be positive, we
will have an
extremely simple model of an open inflationary universe. In the meantime, it
would be nice to have a model where we do not have any problems at all with
synchronization and
with  large fluctuations on the scalar field in the false vacuum. We will
consider such a model in the next section.

\section{\label{Hybrid} Hybrid inflation  with $\Omega < 1$}

 The model to be discussed below \cite{Omega} is a version of the hybrid
inflation scenario
\cite{Hybrid}, which is a slight generalization (and a simplification) of our
previous model
(\ref{3}):
\begin{equation}\label{4a}
V(\phi,\sigma) = {g^2\over 2}\phi^2\sigma^2 + V(\sigma) \ .
\end{equation}
We eliminated the massive term of the field $\phi$ and added explicitly the
interaction ${g^2\over 2}\phi^2\sigma^2$, which, as we have mentioned already,
can be useful (though not necessary)  for stabilization of the state $\sigma =
0$ at large $\phi$. Note
that in this model the line $\sigma = 0$ is a flat direction in the
($\phi,\sigma$) plane. At large $\phi$ the only minimum of the effective
potential with respect to $\sigma$ is at the line $\sigma = 0$.  To give a
particular example, one can take $V(\sigma) = {M^2\over 2} \sigma^2 -{\alpha M
} \sigma^3 + {\lambda\over 4}\sigma^4 +V_0$. Here $V_0$ is a constant which is
added to ensure that $V(\phi,\sigma) = 0$ at the absolute minimum of
$V(\phi,\sigma)$.  In this case the minimum of the potential $V(\phi,\sigma)$
at $\sigma \not = 0$ is deeper than the minimum at $\sigma = 0$ only for $\phi
< \phi_c$, where $\phi_c = {M\over g}\sqrt{{2\alpha^2\over  \lambda} -1}$. This
minimum for $\phi = \phi_c$ appears at $\sigma = \sigma_c = {2\alpha M\over
\lambda}$.

The bubble formation becomes possible only for $\phi < \phi_c$. After the
tunneling the field $\phi$ acquires an effective mass $m = g\sigma$ and begins
to move towards $\phi = 0$, which provides the mechanism for the second stage
of inflation inside the bubble. In this scenario evolution of the scalar field
$\phi$ is exactly synchronized with the evolution of the field $\sigma$, and
the universe inside the bubble appears to be open.

Effective mass of the   field $\phi$ at the minimum of $V(\phi,\sigma)$ with
$\phi = \phi_c$, $\sigma = \sigma_c = {2\alpha M\over  \lambda}$ is   $m =
g\sigma_c = {2g\alpha M\over  \lambda}$. With a decrease of the field $\phi$
its effective mass at the minimum of $V(\phi,\sigma)$ will grow, but not
significantly. For simplicity, we will consider the case $\lambda = \alpha^2$.
 In this case it can be shown that $V(0) = 2.77\, {M^4\over \lambda}$, and the
Hubble constant before the phase transition is given by $4.8\, {M^2\over \sqrt
\lambda M_p}$.  One should check  what is necessary to avoid too large density
perturbations (\ref{o17}). However, one should take into account that the mass
$M$ in (\ref{o17}) corresponds to the  curvature of the effective potential
near $\phi = \phi_c$ rather than at $\phi = 0$. In our case this implies that
one should use $\sqrt 2 M$ instead of $M$ in this equation.  Then one obtains
the following constraint on the mass $M$: \ $M\sqrt \mu   {\
\lower-1.2pt\vbox{\hbox{\rlap{$<$}\lower5pt\vbox{\hbox{$\sim$}}}}\ }
2\times10^{15}$ GeV.  Note that   the thin wall approximation (requiring $\mu
\ll 1$) breaks down  far away from $\phi = \phi_c$. Therefore in general eq.
(\ref{o17}) should be somewhat improved.  However for $\phi \approx \phi_c$ it
works quite well.  To be on a safe side, we will take   $M = 5\times 10^{14}$
GeV.   Other parameters may vary; one may consider, e.g., the theory with  $g
\sim 10^{-5}$, which gives
 $\phi_c = {M\over g}  \sim 5\times 10^{19}\  \mbox{GeV} \sim 4M_{\rm P}$.
The effective mass $m$ after the phase transition is equal to ${2gM\over \sqrt
\lambda}$ at $\phi = \phi_c$, and then it grows by only $25\%$ when the field
$\phi$ changes all the way down from  $\phi_c$ to $\phi = 0$.   As we already
mentioned, in order to obtain the proper amplitude of density perturbations
produced by inflation
inside the bubble one should have $m \sim 10^{13}$ GeV. This corresponds to
$\lambda = \alpha^2 = 10^{-6}$.

The bubble
formation becomes possible only for $\phi < \phi_c$. If it happens in the
interval $4M_{\rm P} > \phi > 3 M_{\rm P}$, we obtain a flat universe. If it
happens at $\phi < 3M_{\rm P}$, we obtain an open universe. Depending on the
initial value of the field $\phi$, we can obtain all possible values of
$\Omega$, from $\Omega = 1$ to $\Omega = 0$. The value of the Hubble constant
at the minimum with $\sigma \not = 0$ at $\phi = 3M_{\rm P}$ in our model does
not differ much from the value of the Hubble constant before the bubble
formation. Therefore we do not expect any specific problems with the large
scale density perturbations in this model.
 Note also that the probability of tunneling at large $\phi$ is very small
since the depth of the minimum at $\phi \sim \phi_c$, $\sigma \sim \sigma_c$
does not differ much from the depth of the minimum at $\sigma = 0$, and there
is no tunneling at all for $\phi > \phi_c$. Therefore
the number of flat universes produced by this mechanism will be strongly
suppressed as compared with the number of open universes, the degree of this
suppression being very sensitive to the value of $\phi_c$. Meanwhile, life of
our type is impossible in empty universes with $\Omega \ll 1$. This may provide
us with a tentative explanation of the small value of $\Omega$ in the context
of our model (see, however, discussion of uncertainties related to this issue
in Sect. \ref{Simplest}).

\section{\label{Natural} ``Supernatural'' inflation with $\Omega < 1$ }

Natural inflation  has been proposed some time ago \cite{BG,Natural} as a
 model in which the inflaton field has a self-coupling constant whose
smallness, required by the amplitude of cosmological inhomogeneities, is
protected by the approximate global symmetry of the underlying particle
physics. The pseudo-Nambu-Goldstone boson (PNGB) field $\phi$, which serves
as an inflaton in these models, would have been exactly massless if not for
explicit
$U(1)$ symmetry breaking induced by non-perturbative effects. The hierarchy
between the scale  of spontaneous symmetry breaking with generation of
Nambu-Goldstone mode and the scale of explicit symmetry breaking which gives a
mass to
this mode is exploited to explain the smallness of the effective mass.

We will consider the case when the PNGB mode is described by the pseudo-scalar
field $\phi$, appearing as a phase of a complex scalar $\Phi$. The scalar
sector
of the effective theory of the pseudo-Nambu-Goldstone mechanism is in general
described by the action:

\begin{equation} \label{axionaction}
S(\Phi) = \int{ d^4x \, \sqrt{-g} \left( g^{\mu \nu} \partial_{\mu} \Phi^*
\partial_{\nu} \Phi
- V_0(|\Phi|) - V_1(\Phi) \right)}.
\end{equation}

In notation $\Phi(x) = {f(x)\over \sqrt 2} \, e^{i{\phi(x) \over f_0}}$ the
field $f(x)$ is the radial component, the PNGB field $\phi(x)$ is the phase
component, and $f_0$ is a dimensional parameter which is equal to the value of
the scalar field $f(x)$ after symmetry breaking.  The function
$V_0(|\Phi|)=V_0(f)$ is the spontaneous symmetry breaking part of the
potential for the complex scalar which remains globally $U(1)$ symmetric under
the transformation $\phi(x) \rightarrow \phi(x) + c$. In the   version
of this theory considered in \cite{Natural} this potential was taken  in the
simplest form
\begin{equation}\label{mexicanhat}
V_0(f) = \frac{\lambda}{4} \left( f^2 - f_0^2 \right)^2 \ .
\end{equation}
The term
 $V_1(\Phi)=V_1(f, \phi)$ is the explicit
$U(1)$ symmetry breaking potential which in many models takes the following
form in the limit $f \rightarrow f_0$:

\begin{equation}\label{explicit}
V_1(f, \phi) = \Lambda^4(f) \, \left(1- \cos\Bigl({\phi \over f_0 } -
\bar{\theta}\Bigr) \right).
\end{equation}

Here $\Lambda^4(f)$ is some relatively slowly varying (in comparison with
$V_0(f)$) function of the radial field, which vanishes at $f=0$.  This term
may appear  due to instantons in a theory with a gauge group with a
``confinement'' scale $\Lambda$, just like the term which is responsible
for the axion mass. Another  reason for appearance of such terms is the
possibility that quantum gravity violates global symmetries
\cite{AW}--\cite{LLKS}. This violation can be described by adding vertex
operators of the type
${g_{nm}\over M_{\rm P}^{m+n-4}}\, (\Phi^n\Phi^{*m}e^{-i\theta(n-m)} + h.c.)$.
After
spontaneous
symmetry breaking the terms with the minimal degree of global symmetry
violation ($|n-m|= 1$) lead to appearance of the terms of the type of
(\ref{explicit}).

The standard assumption of this scenario is that the effective potential
$V_1(f, \phi)$ is much smaller than $V_0(f)$ everywhere except for $f \approx
f_0$. Therefore the functional form of
$\Lambda^4(f)$ should not be very important as far as (\ref{explicit}) is the
leading term in the PNGB effective potential at low energies, but does not give
a
significant contribution into the potential of the radial field away from the
value $f_0$. Thus in what follows we will simply write $\Lambda \equiv
\Lambda(f_0)$ instead of $\Lambda(f)$. Without loss of generality we can assume
$\bar{\theta}=0$ as long
as there are no other terms in the low energy potential depending on it.

The potential in this theory resembles a slightly tilted mexican hat. In
this scenario inflation occurs when the field $\phi$ falls from the
maximum of the potential (\ref{explicit}) at  $\phi \sim \pi f_0$  (for
$\bar{\theta}=0$) to $\phi = 0$. Just
like in the ordinary chaotic inflation scenario, the necessary condition for
inflation to occur is $\pi f_0 {\
\lower-1.2pt\vbox{\hbox{\rlap{$>$}\lower5pt\vbox{\hbox{$\sim$}}}}\ } M_{\rm
P}$.
Inflation will be long enough
and density perturbations produced during this process will have sufficiently
flat spectrum for $f_0 {\
\lower-1.2pt\vbox{\hbox{\rlap{$>$}\lower5pt\vbox{\hbox{$\sim$}}}}\ }  M_{\rm
P}$. For
definiteness, and in agreement with
\cite{Natural}, we will assume here that $f_0 \sim M_{\rm P}$.
The parameter  $\Lambda$ is determined by normalization of density
perturbations produced during inflation.
For $f_0 \sim
M_{pl} \sim 10^{19}$ GeV one must have $\Lambda(f_0)
\sim M_{GUT} \sim 10^{16}$ GeV  (see \cite{Natural} for detailed references).

In order for the radial part of the field to remain frozen and to not
participate in natural inflation there has to be the case that $\lambda >{32
\pi \over 3} \, {\Lambda^4 \over M_{\rm P}^2 f_0^2} \sim 10^{-12}$, which also
ensures that the top of the ``mexican hat'' is higher than the highest energy
of the axion $\Lambda^4$. The coupling constant $\lambda$ could be as large as
unity (however, it does not have to be so large). If we consider the energy
density at the top of the ``mexican hat'' as the measure of how large is the
coupling $\lambda$ (we need this
characterization since we will soon change the shape of the potential), we see
that there is a plenty of room for play --- roughly 12 orders of magnitude in
{\em energy density} between the GUT and Planck scales, or between the scales
of spontaneous and explicit symmetry breaking.

The issue of naturalness of this scenario is not quite trivial. The first time
the possibility of inflation of this kind was investigated by Binetruy and
Gaillard in the context of  superstring theory, and their conclusion was
negative,
for the reason that the typical value of the parameter $f_0$ in superstring
theory is supposed to be few times smaller than $M_{\rm P}$ \cite{BG}. Still,
the
general idea of this scenario is rather elegant. Here we  would like to suggest
its generalization, which would allow us to obtain an inflationary universe
with $\Omega < 1$. We  will call our scenario ``supernatural inflation'' for
its
ability to accommodate naturally low $\Omega$ universes.\footnote{For the
reason mentioned above, we do not want to imply any  relation between this
scenario and supersymmetric theories.}

Our main idea is to construct models which incorporate a primary stage of
``old'' inflation in the false vacuum state near $\Phi = 0$,
and a first order transition which sets up for us the open de Sitter space as a
stage for the subsequent secondary stage of  inflation where the PNGB field
$\phi$ plays the
role of the inflaton. As before, if the number of e-foldings of secondary
inflation will turn out to be just smaller 60, we will find ourselves in an
open, yet matter-rich universe today.

In order to realize this scenario one should have a potential which has a
minimum near $\Phi = 0$. A possible way to achieve it is to add to the
Lagrangian the term $-g^2\chi^2\Phi^*\Phi$
describing interaction of the field $\Phi$  with some other scalar field
$\chi$. If the coupling constant of this interaction is sufficiently large
($g^4
> 16\pi^2 \lambda$), the effective potential of the radial part $f$ of the
field $\Phi$ acquires a new minimum at $f = 0$ due to radiative corrections
\cite{MyBook}.  For $g^4 = 32\pi^2 \lambda$  the minimum of the effective
potential at $f = 0$ becomes as deep as the minimum at $f = f_0$, and the
effective potential  acquires the form $V(f) = {\lambda\over 2} (f_0^2-f^2)f^2
+ \lambda f^4 \ln {f\over f_0}$. Thus for $16\pi^2 \lambda< g^4 < 32\pi^2
\lambda$ the potential has a minimum at $f = 0$ which is somewhat higher than
the minimum at $f= f_0$. If $g^4$ is close to $16\pi^2 \lambda$, the potential
looks like the usual Coleman-Weinberg potential, and the phase
transition from the state $f = 0$ occurs by the Hawking-Moss mechanism.
However, one can show that if $f_0$ is not much greater than $M_{\rm P}$ and if
$g^4$
does not differ much from $32\pi^2 \lambda$ (which means that the minimum at $f
= 0$ is deep enough), then the absolute value of the mass of the field
$f$ always remains greater than the Hubble constant. In such a situation the
phase transition can be well described by the thin wall approximation
neglecting gravitational effects, and (this is important) there will be no
intermediate stage of
inflation during the rolling of the field $f$ towards $f_0$.

We do not want to discuss this issue here in great detail since radiative
corrections is just one of the reasons why the effective potential may acquire
a deep minimum at $f = 0$, and we would like to keep our discussion as general
as possible. In particular, all subsequent results will remain valid if the
potential $V(f)$ is given by the simple expression which we have used in the
previous sections, $V(f) =  {m^2\over 2} f^2 - {\delta\over 3} f^3 +
{\lambda\over 4}f^4$. In order to increase the curvature of the effective
potential at $\Phi = 0$ one may also add the term $\xi R\Phi^2$ to the
Lagrangian. This term, being $U(1)$-invariant,  does not affect the behavior of
the Goldstone mode, and therefore it does not modify the standard picture of
natural inflation, but it changes the curvature of the effective potential. It
also may preclude  inflation at large $f \gg f_0$. This may be useful, since
otherwise inflation may begin at $f \gg f_0$, and then there will  be no first
order phase transition, and the universe will be flat. On the other hand, even
if inflation begins at $f \gg f_0$, one still may experience the second stage
of inflation at $\Phi = 0$ and the subsequent bubble formation if after the end
of this stage of inflation at $f \gg f_0$ the oscillating scalar field  has
enough kinetic energy to climb to the local minimum of the effective potential
at $\Phi = 0$.

An important feature of the ``supernatural'' inflation scenario is a very large
difference between the energy density at the  stage of  inflation at $\Phi =
0$, which has the typical energy scale $\sim {\lambda\over 4} M_{\rm P}^4$, and
the
relatively small energy density $\Lambda^4$ during the last stage of inflation.
As we have already mentioned, these two scales may differ from each other (for
large $\lambda$) by about 12 orders of magnitude.

This implies that after the tunneling there will be a  long intermediate stage
of
non-exponential expansion until the kinetic energy of the radial field and the
energy density of particles produced by its oscillations becomes smaller than
$\Lambda^4$.
 It takes time of the order of  $H^{-1}(\Lambda)\sim M_{\rm P}/\Lambda^2$ to
complete
this intermediate stage. During the sub-luminal expansion epoch $\Omega$ stays
very small
and we can safely assume that the second inflationary stage starts at $\Omega =
0$. One can derive the adiabatic perturbation spectrum, modified by the fact
that the natural inflation starts from the curvature dominated stage.
Typically,
the modification of the density perturbation spectrum is not great (remember,
however, that the observable CMBR temperature fluctuations may differ
considerably, see Section \ref{Bubbles}).

Note that in this scenario the Hubble constant during inflation at $f = 0$ is
much greater than the Hubble constant at the second stage of inflation. In this
respect supernatural inflation resembles the simple model which we discussed in
Sect. \ref{Simplest}. The main difference is that at the first stage of
inflation in the supernatural scenario the mass of the field $f$ is supposed to
be much greater than the Hubble constant. (This condition is satisfied if the
minimum at $f = 0$ is sufficiently deep.) Therefore there are no inflationary
fluctuations of the scalar field produced outside the bubble in the
supernatural inflation scenario. Moreover, even if the mass is not much greater
than $H$, quantum fluctuations at $f = 0$ do not lead to any quantum
fluctuations of the angular field $\phi$, which is responsible for density
perturbations after inflation. Thus, in this scenario we do not have any
complications
related to perturbations penetrating  the bubble from the false vacuum.

In addition to the usual density perturbations produced during the second stage
of inflation inside the bubble, there will be density perturbations induced by
the initial
inhomogeneities of the bubble.   According to eq. (\ref{o14}), we can
estimate the corresponding  density perturbations as follows:
\begin{equation}\label{inducedperturb}
\frac{\delta \rho}{\rho}  \sim
\frac{2\sqrt {2\mu}\Lambda^2}{\sqrt{3\pi} f_0 M_{\rm P}} \sim \frac{2\sqrt
{2\mu}\Lambda^2}{\sqrt{3\pi} M_{\rm P}^2} \ ,
\end{equation}
where the last  approximation assumes $f_0 \sim M_{\rm P}$. For $\Lambda \sim
10^{16}$  GeV and $\mu < 1$ these perturbations are smaller than the usual
perturbations produced during the second stage of
inflation.

If the energy in the maximum of
the radial potential is much greater than the energy of the explicit symmetry
breaking, the tunneling is likely to occur in any direction with equal
probability. If it goes towards $\phi = 0$, one obtains an empty universe with
$\Omega \ll 1$; if it goes towards $\phi \sim \pi f_0$, one obtains a universe
with $\Omega \approx 1$. Thus we may say that it is about as likely to obtain
$\Omega <
1$ as to obtain $\Omega = 1$, if we do not compare the volumes produced during
the secondary inflation (the openness of the universes which we consider makes
comparing the {\em a posteriori} volumes trickier). It is possible,
though, to construct a (fine tuned) model which has a preferred value of
$\Omega$. To do this, we should discover a reason why the phase transition
would go in a given preferred direction rather than any other.

So far we did not consider implications of the global symmetry breaking for the
structure of the potential near $\Phi = 0$. If the corresponding terms appear
due to instanton effects, which become significant only at the late stages of
the universe evolution, the shape of the effective potential at small $f$
remains unchanged.  The physical reason is the infrared cutoff introduced by
the Hawking temperature $T_H = {H\over 2\pi}$ near $\Phi = 0$. (The Hawking
temperature may suppress   effects induced by instantons if  $H(\Phi =
0) \gg \Lambda$.) However, as we already mentioned, in addition to the
low-energy
non-perturbative effects,  the high energy non-perturbative quantum
gravitational effects
may also add symmetry breaking terms of the form  \cite{AW}--\cite{LLKS}
\begin{equation}\label{explicit2}
V_2(\phi(x)) = -{1 \over 2} g_{1}\, M_{\rm P}^3\, (\Phi e^{-i\theta_1} + \Phi^*
e^{i\theta_1}) = - g_{1}\, M_{\rm P}^3\, f(x)
\cos{\left(\frac{\phi(x) }{ f_0 } - {\theta_1}\right)} \ .
\end{equation}

The coupling $g_{1}$ may be strongly suppressed,
 so that  (\ref{explicit2}) does not change the shape of the potential
 (\ref{explicit}) at $f \sim f_0$ (see \cite{LLKS} for
detailed explanations). However the linear term may  play an  important role at
very small $f$, where all other terms are of a higher order in $f$. It
may alter substantially  the shape of the
potential near the top of the mexican hat and
determine the preferable direction of the tunneling. The phase
$\theta_1$, which determines the position of the  minimum of the term
(\ref{explicit2}), does
not have to be the same as $\bar{\theta}$ from (\ref{explicit}), so in general
our potential acquires the form of a  ``twisted'' mexican hat. If $\theta_1$
happens
to coincide with $\bar{\theta}$, then the tunneling typically will produce
empty universes. If $\theta_1$ differs from
$\bar{\theta}$ by $\pi$, we will obtain the bubbles with typical values of
$\Omega \sim 1$. For intermediate values of $\left|\theta_1 -
\bar{\theta}\right|$ we will obtain predominantly $\Omega < 1$ universes. We
should emphasize again that the calculation of probability of creation of
universes with any particular value of $\Omega$ is not unambiguous, see Sect.
5. It is important
 that in all cases the total number of the
universes with any possible value of $\Omega$ will be infinitely large.

\section{\label{Discussion}  Discussion}

In this paper we suggested several different models of a homogeneous
inflationary universe with $\Omega > 1$ and with $\Omega < 1$. At present there
is no observational evidence in favor of $\Omega > 1$. It is clear, however,
that if  observational data are consistent with $\Omega = 1$, they may be
consistent with $\Omega = 1.05$ as well. Therefore it is good to have a working
model of inflationary cosmology with  $\Omega > 1$.

The situation with an open universe may be much more interesting from the point
of view of observational data. That is why in this paper we concentrated on the
discussion of various models of   inflationary universe with $\Omega < 1$.

We have found that in the models containing only  one scalar field one
typically needs fine tuning and rather artificial bending of the effective
potential.
In the models involving two scalar fields one typically obtains infinite number
of open universes with all possible values of $\Omega < 1$ \cite{Omega}. The
simplest model of  this type was discussed in Sect. \ref{Simplest}. This model
is very natural, but it has several unusual features. First of all, the
universe described by this theory is not {\it exactly} homogeneous. A habitable
part of it can be visualized as an exponentially large island inside an
expanding bubble. The difference between the local properties of the
island-like (quasiopen) universe and the homogeneous open inflationary universe
in certain cases becomes very small. In these cases one should carefully
analyse the fate of large perturbations of the scalar field which are generated
outside the bubble but  may penetrate inside it. We gave some arguments
suggesting that these perturbations in this model may be harmless, but this
question requires a more detailed investigation.

In Sect. \ref{Hybrid} we proposed another model \cite{Omega}, which is based on
a certain modification of the hybrid inflation scenario. This model is also
very simple. It describe the universe which is open and homogeneous. Finally,
in Sect. \ref{Natural} we described a modified version of the natural
inflation scenario. We do not want at this
stage to discuss which of the open inflation  models is better. It is clear
that many other models of this type can be proposed.  However, we think that
there is a good chance that many of the qualitatively new  features of the
models discussed above will appear in the new models as well.

Should we take these models seriously? Should we admit that the standard
prediction of inflationary theory that $\Omega = 1$ is not universally valid?
We are afraid that now it
is too late to discuss this question: the jinni is already out of the bottle.
We know that inflationary models describing homogeneous inflationary universes
with $\Omega \not = 1$ do exist, whether we like it or not. It is still true
that the models which lead to
$\Omega = 1$ are much more abundant and, arguably, more natural.  However, in
our opinion, it is
very encouraging that
inflationary  theory is  versatile enough to include models with all
possible values of $\Omega$.

To make our position more clear, we would like to discuss   the history of
the   standard model of electroweak interactions  \cite{GWS}.  Even though this
model was developed by  Glashow, Weinberg and Salam in the 60's, it became
popular only in 1972, when it was realized  that gauge theories with
spontaneous symmetry breaking are renormalizable \cite{Hooft}. However,   it
was immediately pointed out that this model is far from being perfect. In
particular, it  was not based on the simple group of symmetries, and it had
anomalies. Anomalies could destroy the renormalizability, and therefore it was
necessary to invoke a mechanism of their cancellation by enlarging the fermion
sector of the theory. This did not look very natural, and therefore Georgi and
Glashow in 1972 suggested another model \cite{GG}, which at the first glance
looked much better. It was based on the simple group of symmetry $O(3)$, and it
did not have any anomalies. In the beginning it seemed that this model is a
sure winner. However, after the discovery of neutral currents which could not
be described by the  Georgi-Glashow model, everybody forgot about the issues of
naturalness and simplicity and returned back to the  more complicated
Glashow-Weinberg-Salam model, which gradually became the standard model of
electroweak interactions.
This model has about twenty free parameters which so far did not find an
adequate theoretical explanation. Some of these parameters may  appear rather
unnatural. The best example is the coupling constant of the electron to the
Higgs field, which is $2\times 10^{-6}$. It is a pretty unnatural number which
is fine-tuned in such a way as to make the electron 2000 lighter than the
proton.
It is important, however, that all existing versions of the electroweak theory
are based on two fundamental principles: gauge invariance and spontaneous
symmetry breaking.   As
far as these principles hold, we can adjust our parameters and wait until they
get their interpretation in a context of a more general theory. This is the
standard way of development of the elementary particle physics.

For a long time cosmology developed in a somewhat different way, because of the
scarcity of reliable observational data.  Fifteen
years ago many different cosmological models (HDM, CDM, $\Omega = 1$, $\Omega
\ll 1$, etc.) could describe all observational data reasonably well. The main
criterion for a good theory was its beauty and naturalness. Right now it
becomes increasingly complicated to explain all observational data. Therefore
cosmology is gradually becoming a normal experimental science, where the
results of observations play a more important role  than the considerations of
naturalness. However, in our search  for a correct   theory we cannot give up
the requirement of its internal consistency. In particle physics the two
principles which made this theory internally consistent were the gauge
invariance and spontaneous symmetry breaking.  It seems that in cosmology
something like inflation is needed to make the universe large and homogeneous.
It is true that most of the  inflationary models
predict a  universe with $\Omega  = 1$.  Hopefully,  several years later we
will know that our universe is flat, which will be a strong experimental
evidence in favor of  inflationary cosmology in its simplest form. However, if
observational data will show,
beyond any reasonable doubt, that $\Omega \not = 1$, it will not imply  that
inflationary theory is wrong, just like the discovery of neutral currents did
not disprove gauge theories of electroweak interactions. Indeed, now we know
that there is a large class of  internally consistent cosmological models which
may describe
creation of   large homogeneous universes with all possible values of $\Omega$,
and  so far all of these models are based on inflationary cosmology.

The authors are very grateful to  M. Bucher, J. Garc\'{\i}a--Bellido,  L.
Kofman,  V. Mukhanov, I. Tkachev and A. Vilenkin for many enlightening
discussions.  This work was
supported in part  by NSF grant PHY-8612280.

\

\section*{Figure Captions}

Fig. 1  Effective potential $V(\phi, \sigma) = {m^2\over 2}\phi^2 + V(\sigma)$,
eq. (\ref{3}). Arrows show evolution of the fields $\sigma$ and $\phi$. Dashed
lines correspond to tunneling, whereas the solid lines show slow rolling.

\noindent Fig. 2  Effective potential $V(\phi,\sigma) = {g^2\over
2}\phi^2\sigma^2 + V(\sigma)$, eq. (\ref{4a}).

\noindent Fig. 3  Effective potential in the model of supernatural inflation.

\end{document}